\documentclass[english,aps, pra, twocolumn, showpacs]{revtex4}
\usepackage[T1]{fontenc}
\usepackage{verbatim}
\usepackage{amssymb}
\usepackage{graphicx}
\usepackage{epstopdf}
\usepackage{babel}

\makeatletter
\@ifundefined{textcolor}{}
{
 \definecolor{BLACK}{gray}{0}
 \definecolor{WHITE}{gray}{1}
 \definecolor{RED}{rgb}{1,0,0}
 \definecolor{GREEN}{rgb}{0,1,0}
 \definecolor{BLUE}{rgb}{0,0,1}
 \definecolor{CYAN}{cmyk}{1,0,0,0}
 \definecolor{MAGENTA}{cmyk}{0,1,0,0}
 \definecolor{YELLOW}{cmyk}{0,0,1,0}
 }
\makeatother

\begin{document}

\title{Efficient Spin Squeezing with Optimized Pulse Sequences}
\author{C. Shen, L.-M. Duan}
\affiliation{Department of Physics, University of Michigan, Ann Arbor, Michigan 48103,
USA}
\affiliation{Center for Quantum Information, IIIS, Tsinghua University, Beijing 100084,
China}

\begin{abstract}
Spin squeezed states are a class of entangled states of spins that have
practical applications to precision measurements. In recent years spin
squeezing with one-axis twisting (OAT) has been demonstrated experimentally
with spinor BECs with more than $10^{3}$ atoms. Although the noise is below
the standard quantum limit, the OAT scheme cannot reduce the noise down to
the ultimate Heisenberg limit. Here we propose an experimentally feasible
scheme based on optimized quantum control to greatly enhance the performance
of OAT to approach the Heisenberg limit, requiring only an OAT Hamiltonian
and the use of several coherent driving pulses. The scheme is robust against
technical noise and can be readily implemented for spinor BECs or trapped
ions with current technology.
\end{abstract}

\maketitle

Spin squeezed states \cite{ref:Ueda} have attracted a lot of interest due to
both its role in the fundamental study of many-particle entanglement and its
practical application to precision measurements with Ramsey interferometers
\cite%
{ref:wineland,ref:Sorensen_nature,ref:Sorensen_ent_depth,Duan,ref:Toth_entanglement}%
. In recent years, much progress has been made on the experimental squeezing
of a large number ($10^{3}\sim 10^{6}$) of ultracold atoms \cite%
{ref:Takano,ref:Gross_nature,ref:Riedel_nature,ref:vuletic_prl,ref:thompson_prl}%
. Many of these experiments follow the so-called one-axis twisting (OAT)
scheme, which is known to reduce the noise/signal ratio from the classical
case by a amount that scales as $N^{-2/3}$with the particle number $N$ \cite%
{ref:Ueda}. This reduction is not optimal yet and still above the so-called
Heisenberg limit which scales as $N^{-1}$. There have been several
theoretical proposals to enhance the OAT \cite{ref: You_early,ref:LYou}. For
example, one of the approaches \cite{ref: You_early} involves inducing a
better squeezing Hamiltonian, the so called two-axis twisting (TAT)
Hamiltonian, with Raman assisted coupling for trapped spinor BECs. This is a
hardware level engineering, requiring modification of a particular
experimental setup and does not apply to other physical systems. Another
approach \cite{ref:LYou} employs a digital quantum simulation technique to
convert an OAT Hamiltonian to an effective TAT Hamiltonian by
stroboscopically applying a large number of pulses. This software level
solution is universal but sensitive to the accumulation of control errors.
None of these proposals have been experimentally tested yet due to various
difficulties.

Inspired by the idea of optimized quantum control, we propose an
experimentally feasible scheme to greatly improve the performance of OAT,
requiring only two or three additional coherent driving pulses to carry out
collective spin rotations, which is a routine technique with the current
technology. The scheme is shown to be robust to noise and imperfection in
control pulses. Using this scheme, it is possible to generate more spin
squeezing and detect a significantly larger entanglement depth for the
many-particle atomic ensemble \cite{ref:Sorensen_ent_depth}. This new scheme
enhances the OAT squeezing on the software level and therefore can be
applied to any physical system that is endowed with these operations. The
idea of optimized squeezing may also be easily transferred to cases where
the interaction term deviates from the OAT Hamiltonian.

We consider the general scenario of one-axis twisting independent of the
underlying physical system with the Hamiltonian $H=\chi S_{z}^{2}$ ($%
S_{z}=\sum_{i}^{N}s_{z}^{i}$) (setting $\hbar =1$). The system starts from a
collective spin coherent state polarized along $x$-axis. As time goes on the
initially homogenous spin fluctuation gets distorted and redistributed among
different directions and the direction along which spin fluctuation gets
suppressed gradually changes over time. The squeezing is measured by the
parameter $\xi ^{2}$, defined as $\xi ^{2}=N\left\langle S_{\vec{n}%
}^{2}\right\rangle /\left\vert \left\langle S_{x}\right\rangle \right\vert
^{2}$, where $\vec{n}$ is the direction along which spin fluctuation is
minimized. The decreasing rate of $\xi ^{2}$ slows down with time, and after
the optimal squeezing point, $\xi ^{2}$ increases again. Aside from the
initial state, which is rotationally symmetric about $x$-axis, all the
subsequent states breaks this symmetry and picks out a special direction,
i.e. the direction along which fluctuation is minimized. It is well known
that the two-axis twisting (TAT) Hamiltonian  $H_{2}=\chi _{2}\left(
S_{x}^{2}-S_{y}^{2}\right) $ can produce better squeezing \cite{ref:Ueda},
which, after doing the Trotter decomposition with an infinitesimal time
interval, could be seen as switching the squeezing axis back and forth very
fast between two orthogonal directions \cite{ref:LYou}. To avoid the noise
accumulation from a large number of switching pulses inherent in the Trotter
expansion scheme, we take an alternative approach based on optimization of a
few control pulses to maximize the squeezing of the final state. We consider
an $n$-step squeezing protocol (where $n$ is typically $2$ or $3$ for a
practical scheme) defined as follows: at step $j$ ($j=1,\,2,\,...,\,n$), we
first apply an instantaneous collective spin rotation around $x$-axis, $%
U(\alpha _{i})=exp(-i\,S_{x}\alpha _{i})$, and then let the state evolve
under the OAT Hamiltonian $H=\chi S_{z}^{2}$ for a duration $T_{i}$.
Effectively, we squeeze the state along a different axis lying in the $y-z$
plane in each step, so the effective evolution operator can be written as
\begin{equation}
U(\theta _{i},\,T_{i})=\prod_{j=n}^{1}exp(-i\, \chi S_{\theta _{i}}^{2}T_{i}),
\label{eq:evo_operator}
\end{equation}%
where $S_{\theta _{j}}\equiv cos\theta _{j}S_{z}+sin\theta _{j}S_{y}$ and
the factors are arranged from right to left with increase of $j$. Since the
initial state is assumed to be polarized along $x$-direction, which is
symmetric around $x$-axis, $\theta _{1}$ is irrelevant and can be chosen to
be $0$ (so no control pulse is needed for step $1$). Therefore, for an $n$%
-step squeezing protocol, there are $(2n-1)$ tunable parameters: $T_{i}$ and
$\theta _{i}$ (excluding $\theta _{1}$). The final squeezing parameter is
thus a multi-variable function $\xi ^{2}(T_{i},\,\theta _{i})$. Our purpose
is to find the best available squeezing $\xi ^{2}(T_{i},\,\theta _{i})$ with
a minimum number $n$ of the time steps.

\begin{figure}[tbp]
\includegraphics[width=0.4\textwidth]{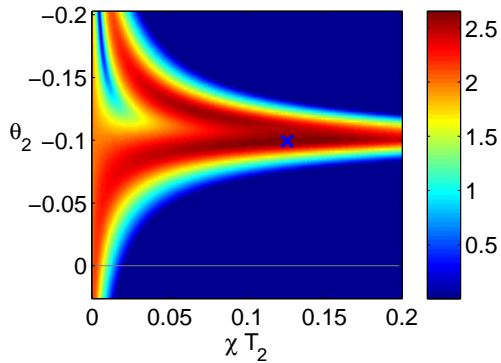}
\caption{The squeezing $-log(\protect\xi ^{2})$ as a function of the control parameters 
$\protect\theta _{2}$ and
$T_{2}$ for a typical value of $T_{1}$, calculated with N=2000 spin-1/2
particles. See Eq. \protect\ref{eq:evo_operator} and text for definition of $%
\protect\theta _{i}$ and $T_{i}$. The cross symbol marks the point of
optimal squeezing. The horizontal line $\protect\theta _{2}=0$ corresponds
to the case of the OAT scheme. }
\label{fig:xi_vs_theta_T}
\end{figure}

In the case of $n=2$ or $3$, the landscape of $\xi ^{2}(T_{i},\,\theta _{i})$
in the parameter space is quite simple and well behaved. Take the $n=2$ case
as an example. For a typical value of $T_{1}$ smaller than the optimal OAT
squeezing time, $-log(\xi ^{2})$ as a function of $\theta _{2}$ and $T_{2}$
is shown in Fig \ref{fig:xi_vs_theta_T}. The optimal squeezing point marked
by the cross lies way off the OAT trajectory, the horizontal line with $%
\theta _{2}=0$. For the $n=3$ case, with $\theta _{2}$ and $T_{2}$ fixed
near the optimal values of the $n=2$ case, $-log(\xi ^{2})$ as a function of
$\theta _{3}$ and $T_{3}$ shows a similar landscape. These solutions already
exceed that of the OAT scheme by a large margin. The results indicate that
the optimization technique with $n$ as small as $2$ or $3$ suffices to
significantly improve over the OAT scheme.

\begin{figure}[tbp]
\includegraphics[width=0.35\textwidth]{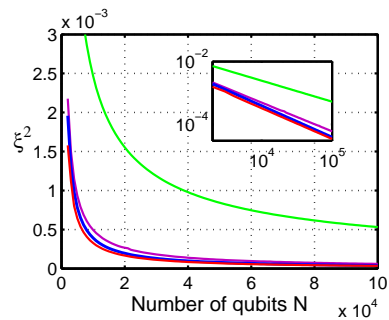}
\caption{Scaling of the squeezing parameter $\protect\xi ^{2}$ with the number of
qubits. Curves from top to bottom are for one-axis twisting (OAT), two-step
optimized squeezing, two-axis twisting (TAT) and three-step optimized
squeezing. Inset shows the same curves in log-log scale. }
\label{fig:xi_vs_N}
\end{figure}

Next, we investigate performance of the optimized squeezing scheme, focusing
on the scaling of the squeezing $\xi ^{2}(T_{i},\,\theta _{i})$ as a
function of the total particle number $N$. For a given set of parameters, we
can numerically calculate the evolution operator in Eq.\ref{eq:evo_operator}
by exactly diagonalizing the effective Hamiltonians $S_{\theta _{i}}^{2}$
and then obtain the squeezing parameter $\xi ^{2}$. We randomly sample from
the parameter space for a large number of times, use these random samples as
initial guesses to start unconstrained local optimization of the squeezing
parameter, and pick the best one as our solution. Repeating this procedure
for every system size $N$ is extremely resource intensive especially when $N$
gets as large as $10^{5}$. Taking advantage of the fact that adding several
more to $10^{3}$ particles should not change the solution much, we can feed
the previously found non-local optimal solution as an initial guess to the
local optimizer of a larger system and obtain a near optimal solution
quickly. In this way we managed to obtain (near) optimal solutions for
systems all the way up to $N=10^{5}$ particles, with only a cost of
classical computing time on the order of tens of hours on a typical
multi-core computer. As shown in Fig \ref{fig:xi_vs_N},  with $n=2$, the
squeezing parameter $\xi ^{2}$ gets reduced by a significant amount already
compared with the OAT scheme, and with $n=3$, $\xi ^{2}$ decreases further.
The scaling of $\xi ^{2}$ with the number of particles shows a clear power
law $\xi ^{2}\sim 1/N^{\beta }$. A simple OAT scheme gives $\beta =2/3$ and
the TAT scheme gives $\beta =1$ \cite{ref:Ueda}. The Heisenberg limit of
noise gives a bound $\beta \leq 1$ for the scaling, and this bound is
saturated by the TAT scheme. Remarkably we observe that the optimized $n=2,3$
protocols can give $\beta =0.92$ and $0.98$, respectively, very close to the
ultimate Heisenberg limit. Moreover, the $n=3$ optimized scheme has a
smaller multiplicative constant compared with the TAT scheme, so in the
realistic range of particle number $N\lesssim 10^{6}$, it actually
outperforms the TAT scheme. This shows that a moderate alternation of the
OAT scheme through optimization can significantly increase the spin
squeezing.

We have demonstrated a significant improvement over the conventional OAT by
applying very few optimized control pulses. A cost of the proposed scheme is
that it takes longer evolution time to achieve the optimal squeezing. A
typical evolution of $\xi ^{2}$ with time $t$ is shown in Fig. \ref%
{fig:xi_vs_time}. We notice that in general the ($i+1$)-th squeezing step
takes longer time than the $i$-th step. Since the time cost in the first
step is on the order of the optimal OAT duration, the overall duration of
the new protocol is usually longer than that of the OAT scheme. An
excessively long duration would be an obstacle in systems with short
coherence time. The two relevant time scales here are the coherence time $%
\tau $ and the inverse of interaction strength $1/\chi $. The time cost of
the new scheme is around $0.01/\chi \sim 0.1/\chi $. If $\tau \gtrsim
0.1/\chi $ the new scheme can be implemented without compromise. On the
other hand, if that is not the case, decoherence effect would play a role
and our unconstrained optimization no longer yields the best result.
However, we can work around this problem by performing an optimization with
the total duration added as a cost function and get a compromised optimal
pulse sequence. By tuning the weight of the cost function we could obtain a
continuous series of compromised optimal solutions as shown in Fig. \ref%
{fig:xi_vs_duration}. These solutions of two-step and three-step schemes
form two line segments, continuously connecting the optimal OAT squeezing
protocol to that of the unconstrained optima, offering a trade off between
the protocol duration and the squeezing magnitude. For each real
experimental setup, one could correspondingly pick up the best point in
accordance with the coherence time of the system. How much one can gain over
the OAT scheme depends on how long the coherence time can reach.

\begin{figure}[tbp]
\includegraphics[width=0.35\textwidth]{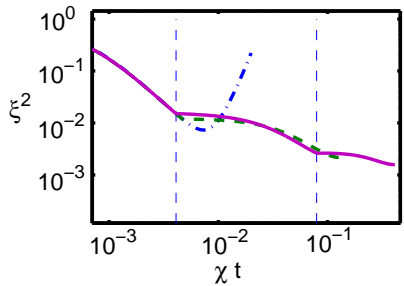}
\caption{Evolution of the squeezing parameter $\protect\xi^{2}$ with time, calculated with N=2000
spin-1/2 particles. The dash-dot line is for one-axis twisting (OAT), 
the dash line for the two-step optimized squeezing scheme, and the solid
line for the three-step optimized squeezing. }
\label{fig:xi_vs_time}
\end{figure}

\begin{figure}[tbp]
\includegraphics[width=0.35%
\textwidth]{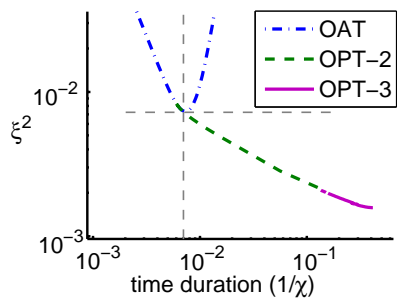}
\caption{Constrained optimization of $\protect\xi^{2}$ with the total time
duration as a cost function. We take $1/\protect\chi$ as the time unit. Achievable
squeezing $\protect\xi^{2}$ as a function of the total duration is shown,
together with one-axis twisting (OAT),
calculated with $N=2000$ spin-$1/2$ particles. OPT-2 (3) stands for optimized
squeezing sequence with $n=2 (3)$ segments. Horizontal and vertical dashed lines
are guides to the eye. }
\label{fig:xi_vs_duration}
\end{figure}

\begin{figure}[tbp]
\centering
\includegraphics[width=0.5\textwidth]{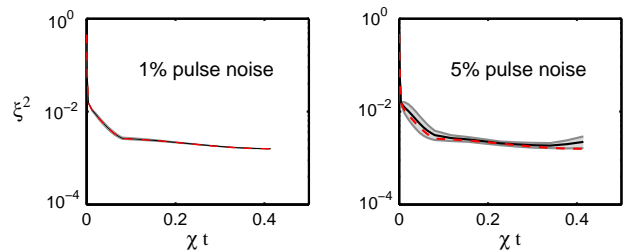}
\caption{Optimized squeezing in the presence of control noise. We use the three-step 
optimization scheme as an example and assume all the five control parameters in this scheme
have the same magnitude of relative errors as specified in this figure. The dash line is for the
ideal case with no error in the control parameters, the solid line denotes the average of many random
trajectories (about $50$ random trials) and the shaded area marks the range of
those trajectories. In the left panel, the shaded region is too small to be
distinguished from the ideal case. }
\label{fig:noisy_control}
\end{figure}

Next we test noise resistance of the proposed scheme. There are only $3(5)$
control parameters in the $n=2(3)$ scheme, making the accumulation of
control noise negligible. We have done numerical simulation of our scheme
adding random pulse area/timing noise and confirmed the robustness of the
squeezing parameter $\xi ^{2}$ as shown in Fig. \ref{fig:noisy_control}.
This contrasts to the proposals \cite{ref:LYou,ref:Jaksch} requiring a large
number of coherent rotation pulses where control errors accumulate and
significantly degrade the performance. Thus our proposed scheme offers a
useful alternative to the previous works. Another practical issue related to
control noise is the uncertainty in number of particles in a real
experiment. Our pulse scheme depends on the number of particles $N$ while in
experiments such as ultracold gas we do not typically know the number $N$
exactly. Fortunately we notice that the control parameters vary slowly with $%
N$ and an uncertainty in $N$ is equivalent to a small extra noise in the
control parameters, to which $\xi ^{2}$ is not so sensitive as we have shown
in Fig. \ref{fig:noisy_control}.

Finally we discuss possible physical realizations of the scheme proposed
here. The scheme only requires two ingredients, the nonlinear collective
spin interaction $S_{z}^{2}$ and the ability to rotate the collective spin
around an orthogonal axis, say $x$. Several experimental systems meet these
requirements, e.g., trapped ions and spinor BECs. In trapped ion systems,
depending on the ion species, one can use bichromatic lasers or two pairs of
Raman laser beams (the Molmer-Sorensen scheme) to induce the $S_{z}^{2}$ or $%
S_{x}^{2}$ type of interaction. The strength of this interaction $\chi $ can
reach kHz scale, giving $1/\chi \sim ms$. The coherence time usually exceeds
$1/\chi $ and our scheme can apply without compromise. Collective spin
rotation can be simply done by shining laser on all the ions driving the
corresponding single-qubit $\sigma _{x/y}$ or rotation. The rotation pulses
have durations much shorter than $1/\chi $. While linear Paul traps \cite%
{ref:blatt_monroe} can now coherently control only about a dozen of ions,
too few for the purpose of spin squeezing, planar Penning traps can
manipulate more than $200$ ions \cite{ref:Bollinger_nature}. For the purpose
of precision measurement, $200$ ions may seem less impressive than $10^{5}$
particles, but we show that using our scheme we can create genuine
multi-particle entangled states with a significantly larger entanglement
depth.  The entanglement depth, defined in \cite{ref:Sorensen_ent_depth}, is
a way to measure how many particles within the whole sample have been
prepared in a genuine entangled state.  Our result is shown in Fig. \ref%
{fig:ent_depth_200}. In this figure, a point lying below the optimal
squeezing curve of $n$ particles correspond to a state that contains genuine
$n$-particle entanglement. Our scheme produces states that lie below the OAT
states in a large range of $\left\langle S_{x}\right\rangle $ values, which
means that experimentally one can achieve a significantly larger
entanglement depth by this optimization technique.

\begin{figure}[tbp]
\includegraphics[width=0.35\textwidth]{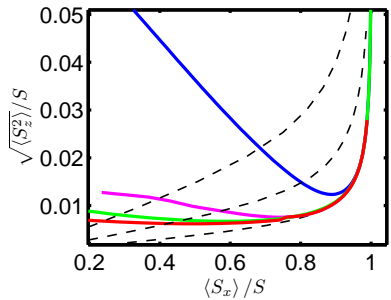}
\caption{The entanglement depth achievable with different approaches for $200$
spin-$1/2$ particles. The solid lines from top to bottom correspond respectively to 
the OAT scheme, the two-step optimized squeezing, the TAT, and the three-step optimized squeezing.
The dashed lines from top to bottom correspond to the optimal squeezing for $50$,
$100$, and $200$ particles respectively. Lying below the curve of optimal
squeezing for $n$ particles is a certificate of genuine $n$-particle
entanglement. }
\label{fig:ent_depth_200}
\end{figure}

Another class of physical system is a spinor Bose-Einstein condensate of
atoms with two chosen internal states mimicking spin-$1/2$ particles \cite%
{ref:Gross_nature,ref:Riedel_nature}. The desired $S_{z}^{2}$ interaction is
induced by spin-dependent s-wave scattering as proposed in \cite%
{ref:Sorensen_nature}. Coherent laser pulses illuminating the whole
condensate can implement spin rotations similar to the trapped ion case.
However, the strength of $S_{z}^{2}$ interaction is much smaller compared
with the trapped ion case, $\chi =0.3\sim 0.5$ Hz as reported in \cite%
{ref:Gross_nature,ref:Riedel_nature}. The coherence time for the spinor BEC
is also shorter. Hence we typically need to apply the compromised scheme,
using the actual coherence time and interaction strength of the system as
input parameters.

In summary, we have proposed a new method based on optimization to
significantly enhance spin squeezing using the one axis twisting
Hamiltonian. To achieve significant improvement in spin squeezing, we need
to apply only one or two global rotation pulses at an appropriate evolution
time and with optimized rotation angles. Using two pulses, the final
squeezing is very close to the Heisenberg limit already. As we use a very
small number of control pulses, the scheme is immune to accumulation of
control errors and can be readily applied in experimental systems without
significant modification of the setup.

This work was supported by the NBRPC (973 Program) 2011CBA00300 (2011CBA00302),
the IARPA MUSIQC program, the ARO and the AFOSR MURI
program, and the DARPA OLE program.

\end{document}